\documentclass[conference]{IEEEtran}
\IEEEoverridecommandlockouts
\usepackage{url}
\usepackage{cite}
\usepackage{amsmath,amssymb,amsfonts}
\usepackage{algorithmic}
\usepackage{booktabs}
\usepackage{graphicx}
\usepackage{textcomp}
\usepackage{xcolor}
\newcommand{\Publicationready}{true}

\def\BibTeX{{\rm B\kern-.05em{\sc i\kern-.025em b}\kern-.08em
    T\kern-.1667em\lower.7ex\hbox{E}\kern-.125emX}}
\begin{document}

\title{Exploring the User Experience of AI-Assisted Sound Searching Systems for Creative Workflows}

% TODO Cars: Bmw 320i Convertible - BMW 320i convertible - drive with top down.

\ifthenelse{\equal{\Publicationready}{true}}{\author{\IEEEauthorblockN{Haohe Liu$^1$, Thomas Deacon$^1$, Wenwu Wang$^1$, Matt Paradis$^2$, Mark D. Plumbley$^1$}
\IEEEauthorblockA{$^1$Centre for Vision, Speech and Signal Processing, University of Surrey, Guildford, UK \\
$^2$Research and Development, British Broadcasting Corporation, London, UK\\
} 
}}
  {\author{\IEEEauthorblockN{Anonymous Authors}}}

\maketitle

\begin{abstract}
Locating the right sound effect efficiently is an important yet challenging topic for audio production. Most current sound-searching systems rely on pre-annotated audio labels created by humans, which can be time-consuming to produce and prone to inaccuracies, limiting the efficiency of audio production. 
Following the recent advancement of contrastive language-audio pre-training~(CLAP) models, we explore an alternative CLAP-based sound-searching system~(CLAP-UI) that does not rely on human annotations. 
To evaluate the effectiveness of CLAP-UI, we conducted comparative experiments with a widely used sound effect searching platform, the \textit{BBC Sound Effect Library}. 
Our study evaluates user performance, cognitive load, and satisfaction through ecologically valid tasks based on professional sound-searching workflows.
% Our experiment evaluates user performance, task load, and satisfaction through tasks resembling real-world sound-searching workflows.
Our result shows that CLAP-UI demonstrated significantly enhanced productivity and reduced frustration while maintaining comparable cognitive demands. 
We also qualitatively analyzed the participants' feedback, which offered valuable perspectives on the design of future AI-assisted sound search systems. 
% These findings highlight the potential of AI-powered tools to transform creative workflows while underscoring the importance of user-centred design and continued innovation in human-AI interaction.
\end{abstract}

\begin{IEEEkeywords}
CLAP, sound effect searching, cognitive load performance
\end{IEEEkeywords}

\section{Introduction}
\label{sec: introduction}
Searching for sound effects~(SFX) is important in multiple sectors, including film, radio, gaming, and interactive media. High-quality sound effects can enhance storytelling and create immersive sensations in audio-visual productions~\cite{playing-with-sound}. 
For example, Kock and Louven~\cite{kock2018power} demonstrated that when participants watched films with sound effects, compared to without sound, their perceived immersion increased more than threefold.
% indicating significantly enhanced viewer immersion.
Furthermore, electroencephalogram~(EEG) studies have demonstrated that well-designed sound effects can positively influence audience engagement, contributing to the formation of general mood and stimulating emotional responses~\cite{kwon2022impact}. 
With the important role of sound effects, sound retrieval also becomes a key part of creative and production workflows.

Traditional sound effects libraries such as Freesound~\cite{font2013freesound} and the BBC Sound Effect Library~(BBC SFX)~\cite{bbc_rewind} have been widely used by audio professionals. BBC SFX contains over $33,000$ sound clips with text annotations, while Freesound allows users to upload sounds with tags for future retrieval. 
Although these libraries have been widely adopted in current audio production workflows, their effectiveness heavily depends on the quality and accuracy of manual annotations, which are time-consuming and labor-intensive. 
For example, an SFX of an engine sound labeled as \textit{``Cars: Bmw 320i Convertible - BMW 320i convertible - drive with top down.''} may fail to capture the primary audio event, making it difficult for users to retrieve this sound using queries like ``engine''.
Besides, the searching systems in most traditional sound effect libraries discourage users from using complex text queries~\cite{baeza2005modeling, weck2024language}. 
% These systems focus on performing retrieval based on word matching, requiring the search query to be locatable in the metadata and users usually do not expect complex queries describing element interactions and temporal orders to work. 
These systems rely on word-matching retrieval, requiring the search query to be directly present in the metadata. As a result, users typically do not expect complex queries involving element interactions or temporal orders to work effectively.
Short query-based searching behavior is potentially not optimal as the user cannot provide detailed text controls for sound retrieval. 
The issue of word matching could be improved with semantic text-to-text matching~\cite{jiang2019semantic} to retrieve a sound with a semantically matched prompt (e.g., retrieve ``barn swallow'' with query ``bird''). 
Text-to-text retrieval still requires manually annotating each audio file. To avoid this manual work, we explore how to match text queries directly with audio content.
% However, text annotation for each audio in the library is still needed to perform text-to-text retrieval. 
% Therefore in this work, we focus on exploring the semantic text-to-audio matching as we would also like to circumvent the need for annotation.
% Therefore, to circumvent the reliance on manual annotation, this work focuses on exploring semantic text-to-audio matching.
% These challenges raise a key question for this paper: \textit{How to design a scalable language-based sound searching system without reliance on human annotations, and how would this system impact the efficiency of audio production?}

Contrastive language-audio pre-training~(CLAP)~\cite{wu2023large-clap} has recently shown success in aligning audio and text modalities in an embedding space with paired audio and text encoders. CLAP is usually trained on a large-scale audio-text paired dataset and learns a joint embedding space where semantically similar audio and text samples are close in distance. 
The aligned latent space enables CLAP to match natural language queries with the nearest audio representations to perform text-based audio retrieval~\cite{elizalde2019cross, 9707629}.
Since CLAP-based sound retrieval doesn't need human annotations, it can handle large audio collections efficiently. Users can search with natural language instead of keywords, making CLAP potentially more intuitive than traditional search methods. We test how well this approach works for professional audio production workflow.
% This retrieval process does not rely on human annotations, so CLAP-based systems can efficiently handle large-scale audio datasets without additional labeling effort. 
% Moreover, the ability to find sounds using natural language makes CLAP potentially more intuitive than traditional sound searching methods, positioning it as a promising candidate for the next generation of sound retrieval systems. This paper explores the impact of CLAP-based sound retrieval systems on audio production performance. 

% To bridge the gap between the CLAP model and sound effect library users,
In this work, we developed a CLAP-based sound searching system~(\textbf{CLAP-UI}) that enables sound searching with natural language. 
% We conducted expert user studies to compare its performance with the word matching based searching system implemented on BBC SFX~\cite{bbc_rewind}, which we refer to as the \textbf{BBC-SFX-UI}. 
% Our research includes two stages. 
We compared CLAP-UI against BBC SFX's word-matching search system [5], which we call \textbf{BBC-SFX-UI}, in a two-stage user study with audio professionals.
For stage one of our study, we developed a prototype sound retrieval system and collected early feedback from participants. This stage focused on gathering user feedback about our prototype system, followed by system improvements. 
For stage two, we recruited additional professional audio producers and designed sound source retrieval tasks to mimic their daily workflows. Specifically, participants were tasked with finding and selecting sounds described in real radio drama scripts.
% , simulating the practical challenges they face in their creative workflow. 
This paper presents the protocol and results from our second stage of testing. We don't discuss the first stage, which gathered feedback on our prototype without comparing systems.
% In this paper, we focus primarily on presenting the experiment protocol and the result of the second stage, as the first stage focuses on the feedback on prototype development and does not perform comparative studies. 

% NASA-TLX is a widely used tool for measuring perceived workload in six dimensions, including mental demand, physical demand, temporal demand, performance, effort, and frustration. For qualitative analysis, we ask open questions such as \textit{What did you dislike most about using the AI-assisted Sound Searching System?} 

To evaluate and compare the performance of CLAP-UI and BBC-SFX-UI, we collected quantitative and qualitative data in our expert user study, including retrieval task performance, general qualitative feedback, and the modified NASA task load index~\cite{hart2006nasa}. 
The results of our study indicate that participants perceived notable benefits when using the proposed CLAP-UI tool compared to the BBC-SFX-UI. 

\section{Study Design and Method}

This section introduces the overall study design, including the background of CLAP~(Section~\ref{sec: clap}), the design of our retrieval system~(Section~\ref{sec: user-interface-design}), the protocol of the expert user study~(Section~\ref{sec: participatory-study}), and the participant background analysis~(Section~\ref{sec:participant-background}). 

\subsection{CLAP-based Audio Retrieval}
\label{sec: clap}
% Contrastive Language-Audio Pretraining (CLAP)~\cite{wu2023large-clap, elizalde2023clap} is a state-of-the-art framework designed to align audio and textual modalities within a shared latent space. This section introduce the model architecture, optimization .

\subsubsection{Model Architecture}

The CLAP~\cite{wu2023large-clap} model primarily consists of an audio encoder and a text encoder. 
% The audio encoder $f_{\text{audio}}(\cdot)$ processes input audio data $X^a$ to generate embeddings $E^a \in \mathbb{R}^D$. 
% The text encoder $f_{\text{text}}(\cdot)$ processes input text data $X^t$ to produce embeddings $E^t \in \mathbb{R}^D$. 
The audio encoder \( f_{\text{audio}}(\cdot) \) and text encoder \( f_{\text{text}}(\cdot) \) process input audio \( X^a \) and text \( X^t \), respectively, to generate embeddings \( E^a, E^t \in \mathbb{R}^D \), where $D$ is the dimension of the embeddings.
Popular options for the text encoder include BERT~\cite{devlin2018bert}, RoBERTa~\cite{RoBERTa}, and BART~\cite{lewis2019bart}. 
The text encoder output is further mapped to the shared embedding space using a two-layer multilayer perceptron (MLP).
Common architectures for the audio encoder include the pretrained audio neural networks (PANNs)~\cite{kong2020panns}, and the hierarchical token-semantic audio transformer (HTSAT)~\cite{HTSAT}. PANNs is a convolutional neural network (CNN) based model with downsampling and upsampling blocks, initially designed for audio classification, while HTSAT is a transformer-based model with hierarchical token processing.
% Common architectures for the audio encoder include the pretrained audio neural networks~(PANNs), a convolutional neural network (CNN) based model with downsampling and upsampling blocks, initially designed for audio classification, and the hierarchical token-semantic audio transformer~(HTSAT), a transformer-based model with hierarchical token processing. 
In a similar way to the text embeddings, the audio embeddings are mapped to the shared embedding space using a two-layer MLP. 
The calculation of audio and text embeddings can be formulated as follows:
\begin{equation}
    E^a = \text{MLP}_{\text{audio}}(f_{\text{audio}}(X^a)), \quad
E^t = \text{MLP}_{\text{text}}(f_{\text{text}}(X^t)).
\end{equation}
In this work, we adopt the pre-trained CLAP from~\cite{wu2023large-clap}, developed based on RoBERTa and HTSAT. 

\subsubsection{Optimization}

The audio and text encoders are trained using a contrastive loss function to align the audio and text embeddings. The loss ensures that paired audio-text samples are mapped closer in the shared space while non-paired samples are pushed further apart. The contrastive loss is given by
\begin{equation}
L = \frac{1}{2N} \sum_{i=1}^N \left( \log \frac{e^{E^a_i \cdot E^t_i / \tau}}{\sum_{j=1}^N e^{E^a_i \cdot E^t_j / \tau}} + \log \frac{e^{E^t_i \cdot E^a_i / \tau}}{\sum_{j=1}^N e^{E^t_i \cdot E^a_j / \tau}} \right),
% \begin{split}
%     L = \frac{1}{2N} \sum_{i=1}^N \Bigg( & \log \frac{\exp(E^a_i \cdot E^t_i / \tau)}{\sum_{j=1}^N \exp(E^a_i \cdot E^t_j / \tau)} \\
%     & + \log \frac{\exp(E^t_i \cdot E^a_i / \tau)}{\sum_{j=1}^N \exp(E^t_i \cdot E^a_j / \tau)} \Bigg),
% \end{split}
\end{equation}

\noindent
where $\tau$ is a learnable temperature parameter, and $N$ is the batch size. The two logarithmic terms represent audio-to-text and text-to-audio alignment, respectively.

\subsubsection{Text-to-Audio Retrieval}

Once pre-trained, the CLAP model can perform text-to-audio retrieval by projecting text and audio into the shared embedding space and computing their similarity. For a given text query $X^t$, the text embedding $E^t$ is computed using the text encoder. The system then identifies the audio embedding $E^a$ in the dataset that maximizes the following cosine similarity
\begin{equation}
    \text{Similarity}(E^t, E^a) = \frac{E^t \cdot E^a}{\|E^t\| \|E^a\|}.
\end{equation}
Given a text query, the retrieval result is a list of audio clips ranked by similarity score. 
% This approach ensures efficient and scalable search in large audio databases without the need for manually curated annotations.

\subsection{User Interface Design}
\label{sec: user-interface-design}

As noted in Section~\ref{sec: introduction}, the retrieval system is developed in two phases. The first stage focused on building a prototype and collecting early feedback to refine the design. In this paper, we concentrate on presenting the final system from the second stage.
% while also discussing the feedback from the first stage to explain the reason behind our interface design choices.

\begin{figure}[tbp]
    \centering
    \includegraphics[width=1.0\linewidth]{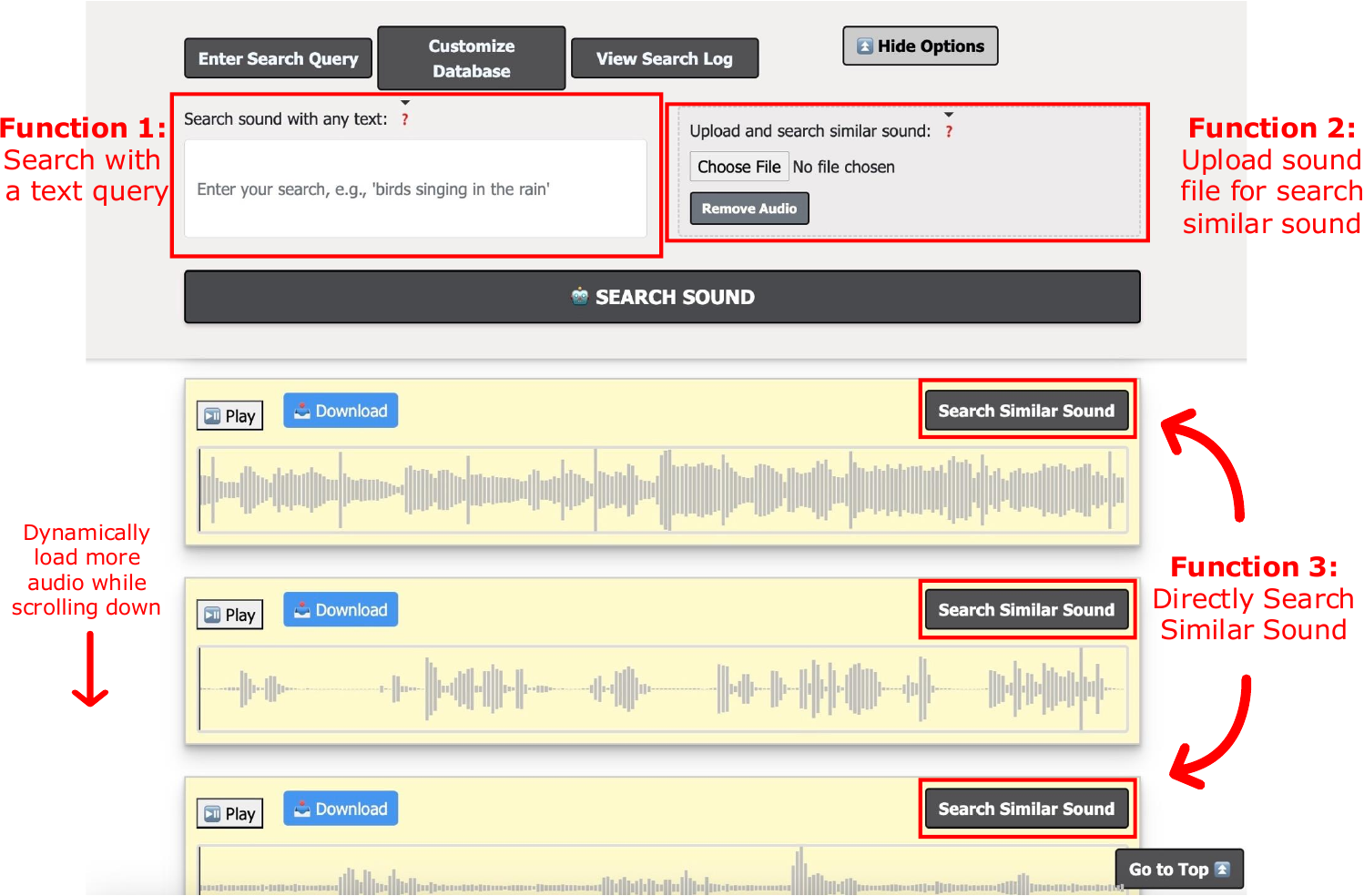}
    \caption{The user interface of our CLAP-based sound searching system.}
    \label{fig:user-interface-CLAP-UI}
\end{figure}

We developed our CLAP-UI as a web application.
% , enabling users to interact with a web browser on any device. 
As shown in Figure~\ref{fig:user-interface-CLAP-UI}, the system offers three core functionalities: (1) searching with a text query, (2) uploading a sound file for search, and (3) utilizing a ``search similar sound'' function. These options can be used independently or in combination, allowing users to provide multiple inputs to enhance search precision. When multiple queries, such as text and audio files, are used together, the system computes the final query embedding by averaging multiple CLAP embeddings. 
The second and third functionalities were added in response to feedback from the prototype development stage, where participants reported challenges in describing sound requirements solely through text. To address this, we designed multimodal search capabilities that combine textual and audio-based inputs.

The audio database used in CLAP-UI is AudioSet~\cite{gemmeke2017audioset}, comprising $1,912,134$ 10-second audio clips labeled across $527$ classes. During the prototype stage, feedback emphasized the importance of system responsiveness. Based on a widely used Python web demo framework Gradio~\cite{abid2019gradio}, our initial prototype required more than $6$ seconds to render the results. To enhance the system responsiveness, we re-developed the system with javascript and Flask~\cite{grinberg2018flask} with searching algorithm optimizations. As a result, the final system achieves an average response time of under $0.5$ seconds for a dataset containing nearly two million audio files. Additional features, such as unlimited scrolling and database customization, are also implemented to enhance the user experience. However, for our experiment, users were restricted to searching within AudioSet to ensure consistency in experimental conditions.

% The design of these features was informed by feedback collected during the first stage of prototype development, where participants identified key challenges such as difficulty in expressing sound needs through text alone and uncertainty about whether their desired sound existed in the dataset. For example, some users noted, \textit{Sometimes it is hard to conceptualize the sound that I need in words to produce a search prompt}.
% and suggested it would be useful to \textit{understand whether the target sound is not in the dataset at all or the system cannot find a good match}. 
% To address these concerns, we introduced multimodal search inputs and ensured that users could combine text and sound queries to increase search precision. Besides, feedback from participants highlighted the importance of system responsiveness, which is around $6$-second for our prototype system. To address this response speed issue we re-implement our system with a more efficient framework and do performance optimization. 

% Response time
% Deployment
% CLAP checkpoint

\subsection{Expert User Study Design}
\label{sec: participatory-study}

The experimental procedure was conducted fully online, with participants completing tasks and providing feedback remotely. Communication was facilitated with email, which served as the primary method for sharing instructions, addressing questions, and following up after participation.

% Our study primarily evaluates the impact of a text-based audio retrieval system on the productivity, efficiency, and cognitive workload of professional audio producers. Through structured tasks, questionnaires, and qualitative feedback, our experiment compares the proposed system with the BBC Sound Effects Library. 
% The aim is to measure not only the quantitative performance of the systems but also the subjective experience and perceived usability of the participants.

\subsubsection{Participants and Recruitment}

The study recruited participants with both professional experience in audio production and researchers in the multimedia domain. A total of $38$ individuals expressed interest in participating, with $20$ finally having completed the study. The study was conducted remotely with clear instructions provided via Microsoft Forms\footnote{Example: \url{https://github.com/unkown-me/CLAP-UI-VS-BBC-UI}}. Recruitment was done through professional email networks within 
\ifthenelse{\equal{\Publicationready}{true}}
  {the BBC and the Centre for Vision, Speech, and Signal Processing (CVSSP) at the University of Surrey.}
  {the [ANONYMOUS – Radio Broadcaster] and [ANONYMOUS - Host Institution].} %  within the BBC and the [ANONYMOUS - HOST INSITUTION]
Participants were given a detailed information sheet outlining the purpose of the study, procedures, and ethical considerations. Informed consent was obtained before their participation, ensuring adherence to ethical standards. To acknowledge their time and effort, all participants received reasonable compensation.

\subsubsection{Task Design}
\label{sec:task-design}
\begin{table}[t!]
\centering
\caption{Scripts for the \textit{Sound Source Retrieval} Tasks}
\label{tab:audio_prompts}
\begin{tabular}{c p{7cm}} % Adjusted column width
\hline
\textbf{Script ID} & \textbf{Description} \\ \hline
S-$1$ & Opens a cupboard. \\ \hline
S-$2$ & Sound of an old wooden rowing boat in a still sea. \\ \hline
S-$3$ & A fire is burning in a stove. A man breaks a piece of wood in two and puts it in the stove. \\ \hline
S-$4$ & Distant bells. \\ \hline
S-$5$ & Coin goes into 1970’s phone box. We hear dialling. \\ \hline
S-$6$ & Jazz piano. Footsteps walk on stage. \\ \hline
S-$7$ & The wedding night, in the bed chamber. The heavy oak door slams shut. \\ \hline
S-$8$ & The drip of water in an echoing stone space. The slight murmur of crowds and music far away. \\ \hline
S-$9$ & Sound of taxi pulling up outside a farm. \\ \hline
S-$10$ & A man starts shouting excitedly. \\ \hline
S-$11$ & Walks towards the lifts and presses the button. \\ \hline
S-$12$ & Background: phones, typewriters. \\ \hline
\end{tabular}
\end{table}

The experiment was centered on a ``Sound Source Retrieval'' task designed to simulate real-world audio production scenarios. Participants are instructed to search for sound effects for pre-defined scripts, such as \textit{Zoo ambience with cheering, children laughing, and people talking}, which were sourced from existing radio drama scripts to reflect real-world production settings. 
The $12$ scripts used in the experiment that indicate target sounds are detailed in Table~\ref{tab:audio_prompts}. Each participant performed the task using both the BBC-SFX-UI and CLAP-UI. For each system, participants reviewed the textual description of each script, searched for the most appropriate sound effect, and rated the difficulty of finding a relevant sound on a $0$ to $10$ scale, where $0$ represented \textit{Very easy to find suitable sounds} and 10 represented \textit{Extremely hard to find suitable sounds}. 
% After completing the $12$ scripts for a system, we use a modified NASA task load index~\cite{hart2006nasa} to evaluate the participant's task load index across five dimensions. The modified NASA TLX is a modification and simplication of original NASA TLX~\cite{hart1988development}. 
After completing the $12$ scripts for each system, participants’ task load was assessed using a modified NASA task load index~\cite{hart2006nasa}. This modified version simplifies the original NASA TLX~\cite{hart1988development} by removing the weighting process. Instead, the overall workload is calculated as the average of individual ratings. Additionally, explanations were provided for each dimension to reduce confusion, as detailed in Table~\ref{tab:nasa_tlx_wording}.
% Specifically, we drop the weighting process in the original NASA TLX, the overall workload is simply the average of individual ratings, and add explanations on each dimension to minimize confusion, as shown in Table~\ref{tab:nasa_tlx_wording}.
To minimize bias and ensure fairness, the order of system usage is balanced, with $10$ participants beginning with the BBC-SFX-UI and the other $10$ participants starting with the proposed CLAP-UI system. The same set of sound effect scripts was used for both systems to ensure consistency in evaluation.

\begin{table}[tbp]
\centering
\caption{Wording used for task load evaluations.}
\begin{tabular}{lp{5cm}}
\hline
\textbf{Dimension}       & \textbf{Wording Used in the Questionnaire} \\ \hline
Mental Demand            & How easy or demanding, simple or complex was the task? \\ \hline
Temporal Demand          & How much time pressure did you feel in performing the task? How hurried or rushed was the pace of the task? \\ \hline
Performance              & How successful were you in accomplishing what you were asked to do? \\ \hline
Effort                   & How hard did you have to work to accomplish your level of performance? \\ \hline
Frustration              & How insecure, discouraged, irritated, stressed, and annoyed were you during the task? \\ \hline
\end{tabular}
\label{tab:nasa_tlx_wording}
\end{table}

\subsubsection{Post-task Surveys}
After completing tasks with both systems, participants completed a post-task survey to provide both quantitative and qualitative feedback. Quantitative questions asked participants to rate their experiences on a Likert scale.
% , where 0 represented \textit{Much worse} or \textit{Not at all likely}, and 10 represented \textit{Significantly better} or \textit{Extremely likely}. 
For instance, participants were asked, \textit{Would you consider using the AI-assisted Sound Searching System in your workflow?~(Q1)} and \textit{How well did you perform the task with the AI-assisted Sound Searching System compared to the BBC Sound Effect Library?~(Q2)} Optional comment boxes accompanied these questions, allowing participants to elaborate on their ratings. The average completion time of the questionnaire, including the sound source retrieval task and the post-task survey, is $47$ minutes $14$ seconds.

The qualitative section explored detailed impressions of the AI-assisted system, with questions such as \textit{What did you like~(Q3)/dislike~(Q4) most about using the AI-assisted Sound Searching System?} Participants were also asked to reflect on scenarios where the CLAP-UI performed better or worse than the BBC-SFX-UI. Additionally, demographic information was collected, including prior experience with the BBC-SFX-UI, frequency of sound library usage, educational background, age, and gender. This data enriched the analysis, providing context for participant feedback. 
\ifthenelse{\equal{\Publicationready}{true}}
  {This study received a Favourable Ethical Opinion (FEO) from the University of Surrey Ethics Committee and the Research Integrity and Governance Office (Reference Number: FEPS 22-23 016 EGA) on 27 June 2023. All procedures performed were in accordance with the Committee’s guidelines, and participants provided written informed consent prior to participation.}
  {This study received a Favourable Ethical Opinion (FEO) from the [ANONYMOUS - Host Institution] (Reference Number: [ANONYMOUS]) on 27 June 2023. All procedures performed were in accordance with the Committee’s guidelines, and participants provided written informed consent prior to participation.}

\subsection{Participant Background}
\label{sec:participant-background}
% The study recruited participants from diverse and balanced backgrounds. 
A total of $20$ participants were recruited, with a gender distribution of $55\%$ male and $45\%$ female. Age-wise, $70\%$ of participants are within the $25-34$ age group. The participants are well-educated, with $35\%$ holding master's degrees, $30\%$ with doctoral degrees, and the remaining $35\%$ with a bachelor's degree. $60\%$ of the participants reported previous experience with the BBC-SFX-UI. As for general sound effect libraries, most participants reported prior experience.
% , with only one participant indicating no prior interaction. 

\section{Result Analysis}

This section compares the AI-assisted system (CLAP-UI) with the BBC-SFX-UI regarding the difficulty of the Sound Source Retrieval task. This section analyzes both qualitative feedback from participants and quantitative results from script difficulty ratings and the task load index. For clarity, participants and scripts are referred to as P-$n$ and S-$n$, respectively, where $n$ is the identifier.

\subsection{Analysis Method}
% We used the Wilcoxon signed-rank tests to analyze overall differences between the UI systems for task difficulty ratings. Task load index use similar Wilcoxon signed-rank tests as evaluating the overall difference. 
Due to the ordinal nature of difficulty ratings and repeated measures design, non-parametric statistics are used~\cite{wobbrock2016nonparametric}.
We used the Wilcoxon signed-rank test to evaluate overall differences in task difficulty ratings between the UI systems, with alpha level $\alpha=0.05$. The five dimensions of the modified NASA TLX evaluation were analyzed using the same test. 
For examining specific script-level differences between BBC-SFX-UI and CLAP-UI, Bonferroni corrections were used with Wilcoxon signed-rank tests to reduce the family-wise error rate. As we have 12 prompts in total, the Bonferroni-corrected alpha level becomes $\alpha_{\text{b}}=\alpha/12\approx 0.0042$.
Additionally, we report the effect size~($r$) for each significance test to provide further context and interpret the magnitude of the observed differences. As we have a relatively small sample size, we calculated the effect size with the rank-biserial correlation~\cite{cureton1956rank}, which measures the proportion of the rank sum difference relative to the total rank sum.
% We also report the effective size of each significance test to provide more context on the difference.

\subsection{Task Performance}

\begin{figure}[tbp]
    \centering
    \includegraphics[width=1.0\linewidth]{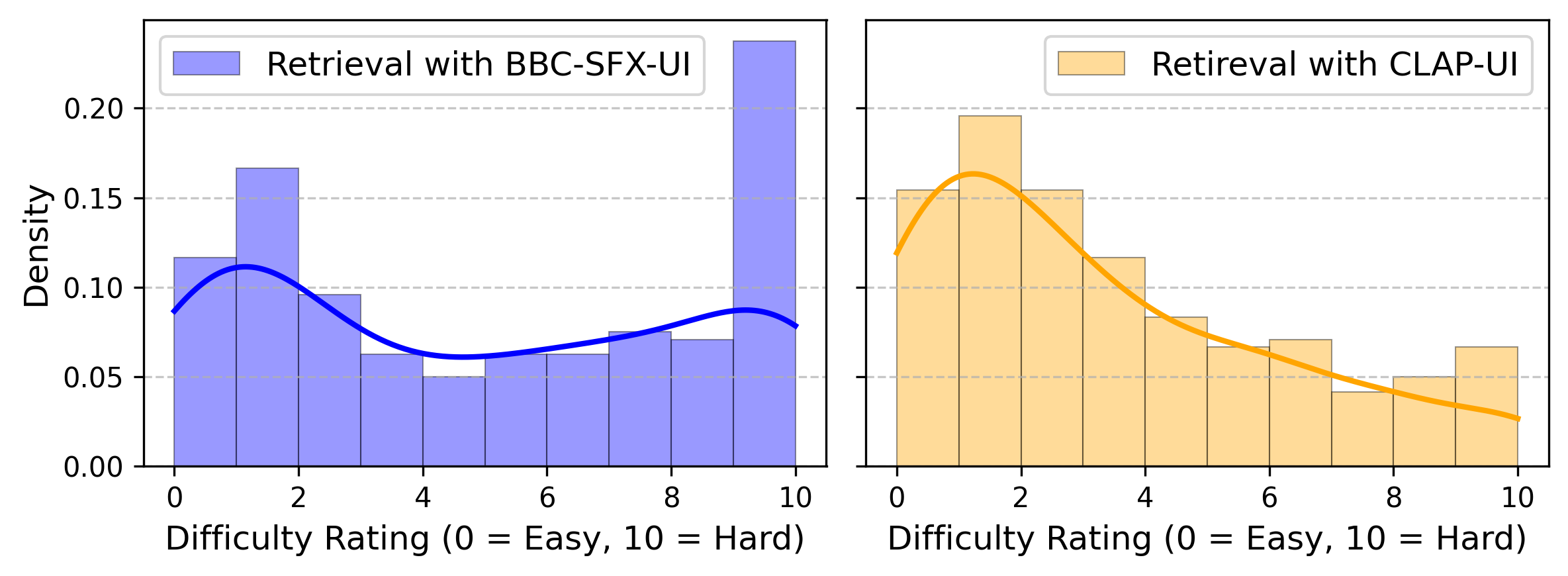}
    \caption{Overall script difficulty distribution (\(p=1.14 \times 10^{-8}, r=0.416\)).}
    \vspace{-1em}
    \label{fig:prompt-difficulty-distribution}
\end{figure}

\begin{figure*}[tbp]
    \centering % Center the figure on the page
    \includegraphics[width=1.0\linewidth]{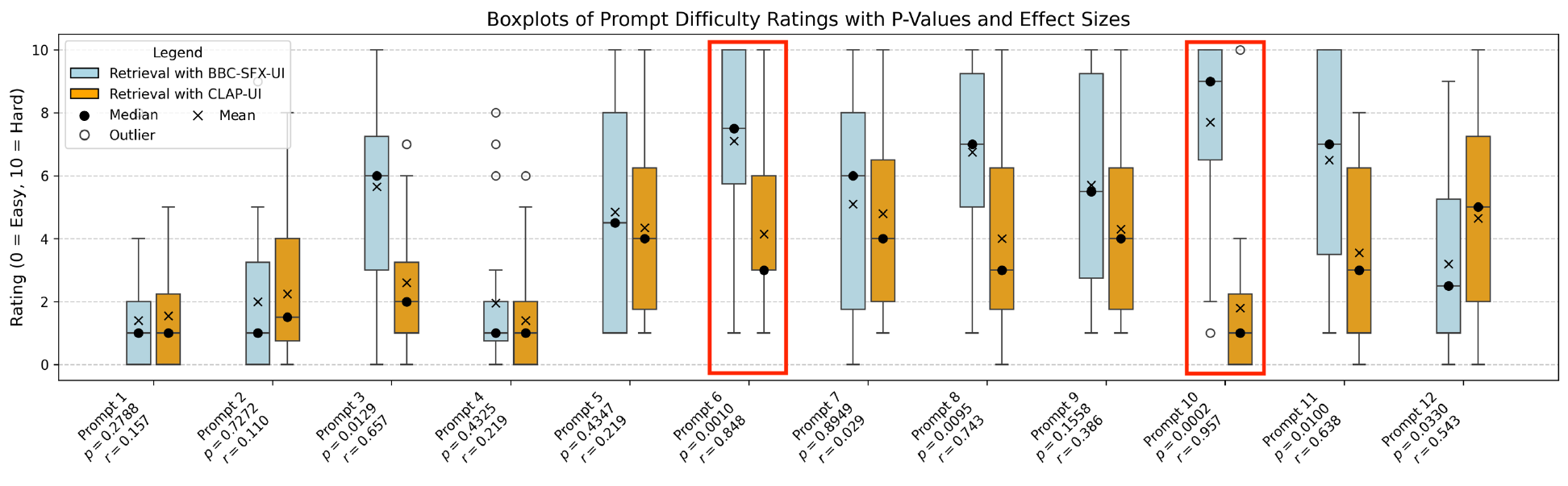}
    \caption{Difficulty rating for each of the sound effect scripts. Red boxes mark the statistically significant results after the Bonferroni correction.}
    \vspace{-1em}
    \label{fig:prompt-score} 
\end{figure*}

The CLAP-UI system demonstrated clear advantages over the BBC-SFX-UI in task performance. Participants consistently found it easier to locate relevant sounds with CLAP-UI, as reflected in significantly lower overall difficulty ratings ($p = 1.14 \times 10^{-8}$, $r = 0.416$). This trend is particularly evident in two specific prompts, S-$6$ and S-$10$, where CLAP-UI significantly outperformed BBC-SFX-UI ($p = 0.001, r=0.848$ and $p = 0.0002, 0.957$, respectively). Figure~\ref{fig:prompt-difficulty-distribution} and Figure~\ref{fig:prompt-score} provide insights into the overall difficulty ratings and each script, respectively. 

\subsubsection{Individual Scripts Difficulty Analysis} Statistical significance after Bonferroni correction (\(p < \alpha_\text{b}\)) is observed for S-$6$~(\textit{``Jazz piano. Footsteps walk on stage''}, $p=0.001, r=0.848$) and S-$10$~(\textit{``A man starts shouting excitedly''}, $p=0.0002, r=0.957$), where the AI system demonstrated better performance compared to the BBC-SFX-UI. In contrast, no statistically significant differences were observed for the rest of the scripts (\(p > \alpha_\text{b}\)).
To look into the statistical significance we got on S-$6$ and S-$10$, we compare the search result of these two prompt on BBC-SFX-UI and CLAP-UI. The top result when searching for \textit{``Jazz piano''} and \textit{``shouting excitedly''} on BBC-SFX-UI are labelled with \textit{``Pianos: Comedy - One piano dragged along''} and \textit{``Animals - Airedale panting excitedly.''}, respectively, which have low relevance. However, CLAP-UI can give exactly the audio required in the top result.

 % S-$10$ showed the largest gap, as participants noted that CLAP-UI effectively retrieved human shouting sounds, whereas BBC-SFX frequently returned irrelevant results, such as animal sounds.

 \subsubsection{Overall Script Difficulty Analysis} 
Figure~\ref{fig:prompt-difficulty-distribution} provides further evidence of the comparative advantage of CLAP-UI (Proposed) over BBC-SFX-UI. The histogram and kernel density estimate~(KDE)~\cite{chen2017tutorial-kde} plots show that difficulty ratings for CLAP-UI are skewed toward lower values, indicating that participants found it easier to locate relevant sounds overall. In contrast, difficulty ratings for the BBC-SFX-UI are more evenly distributed. To statistically validate these observations, we conducted a Wilcoxon signed-rank test, as the data is paired and non-normally distributed, confirmed by the Shapiro-Wilk test (\(p < 0.05\)). The test result (\(p=1.14 \times 10^{-8}, r=0.416\)) demonstrates a significant difference between the two systems and a moderate effect size. 

% These results suggest that the AI-assisted system's semantic understanding and natural language processing capabilities significantly reduce task difficulty, particularly for complex or descriptive prompts, compared to the keyword-based approach of the BBC library.

% In summary, while both systems performed adequately for straightforward prompts, the AI-assisted system significantly reduced difficulty for more nuanced or creative queries. These findings highlight the potential of the AI system to enhance efficiency and user satisfaction in sound retrieval tasks, particularly in scenarios requiring semantic interpretation or handling of multi-element prompts.
% \subsubsection{Overall Impressions and Comparative Performance}

\begin{figure}[tbp]
    \centering
    \includegraphics[width=1.0\linewidth]{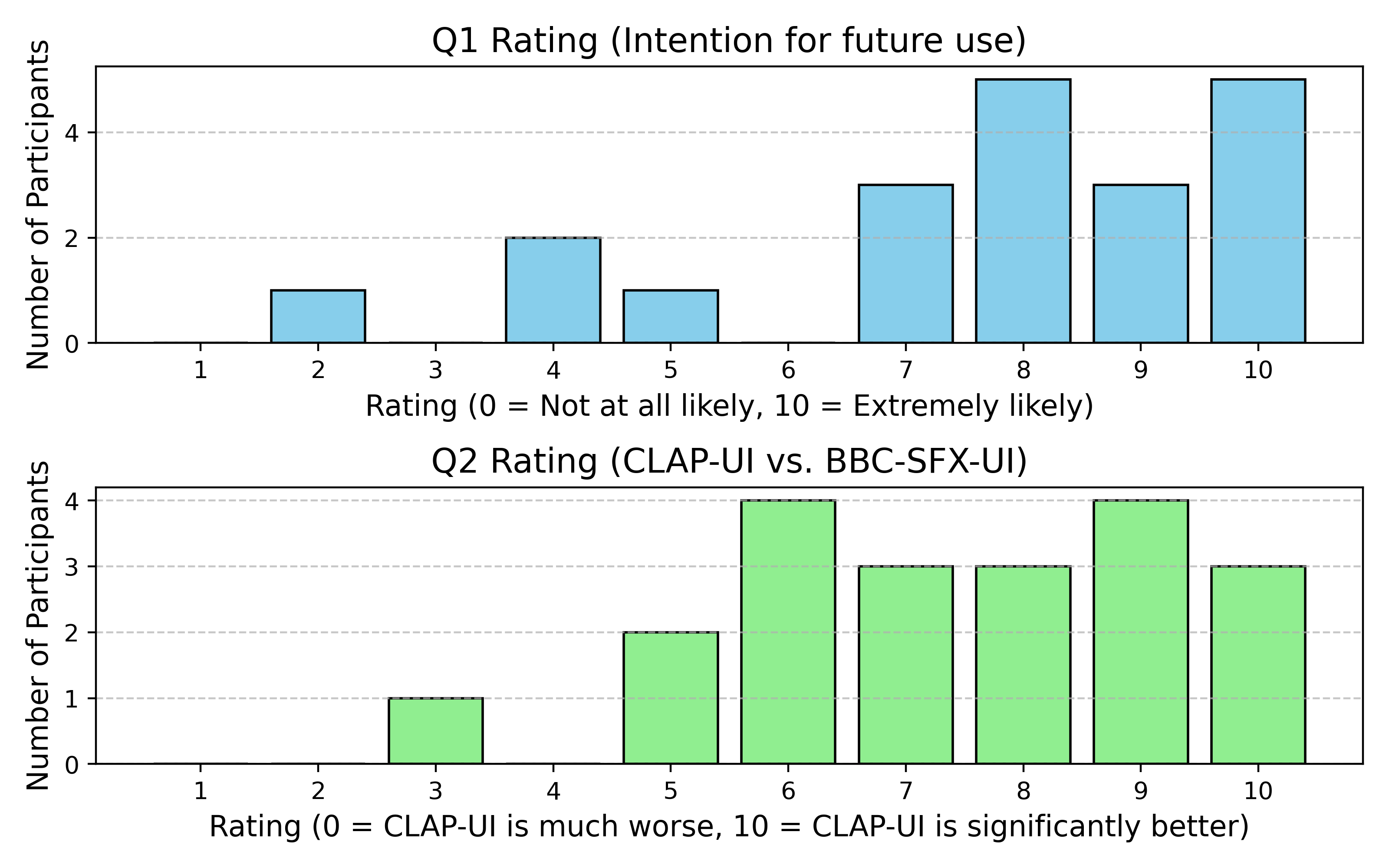}
    \caption{Participant ratings for Q1 and Q2. }
    \vspace{-1em}
    \label{fig:q1-rating}
\end{figure}

\subsection{Preference Ratings and Qualitative Feedback}

As shown in Figure~\ref{fig:q1-rating}, participants provided a range of ratings for Q1 and Q2, with average scores of $7.65$ and $7.4$, respectively. Most participants rate in the mid-to-high range ($6$–$10$), indicating a general preference for the usability of CLAP-UI and a likelihood of adopting it for future use. For instance, P-$19$~(Q1: $10$, Q2: $10$) gives CLAP-UI a score of $10$ for both questions and states \textit{``I can always find audio that I want to find ... The AI-assisted system is easy to use and always works for me.''}

Many participants highlighted the improved relevance of CLAP-UI results compared to the BBC-SFX-UI. For example, P-$10$~(Q1: $9$, Q2: $9$) commented \textit{``Although the AI option struggled with a couple of searches, I always felt like the results indicated an understanding of approximately what I was trying to achieve. The BBC search, time and time again, gave results that were hugely irrelevant (e.g., when searching for a man shouting excitedly [S-$10$], BBC gave me a full page of horse noises). Most impressive was the few occasions when the top result from the AI search was almost exactly what I had in mind!''} 
Similarly, P-$18$ (Q1: $8$, Q2: $7$) notes \textit{``I think the AI system gave more relevant search results. The BBC system, for example, when searching for `Walks towards the lifts and presses the button,' returned top results like bird and water sounds, which were completely irrelevant.''}

Another commonly mentioned benefit of CLAP-UI is efficiency. Several participants described how the AI system reduced the mental overhead of searching for sounds. P-$8$ stated, \textit{``Felt I could get there much faster with the AI-assisted. ... I can only imagine with an even bigger sample library it will only become stronger.''} Similarly, P-$9$ commented, \textit{``Yes, I can generate multiple audio clips from similar scenes using a few keywords quickly and efficiently.''}

The creative flexibility offered by CLAP-UI is another area of strength. Participants appreciated how CLAP-UI allowed them to explore sound design conceptually. P-$1$ observed that the AI-assisted system \textit{``allows me to be more creative in the prompts I’m asking for,''} while P-$17$ highlighted how the system increased creative possibilities by enabling searches for sounds that might not exist in real life, stating, \textit{``[CLAP-UI] can help me find some sounds that you imagine, which may not be relevant or possible in real life, increasing my creative ability.''}

The ``Search Similar Sound'' feature was also well-received, enabling users to refine their results more effectively. P-$9$ emphasized its utility in research requiring multiple sounds with similar environmental atmospheres: \textit{``The search similar sound function is very useful for me, because my research requires video-level semantic relationship learning using multiple sounds.''} 
% This functionality was described as particularly beneficial for narrowing down results to match specific criteria.

% \subsubsection{Drawbacks and Limitations of the AI-Assisted System}

While the AI system demonstrated clear advantages, several limitations were noted. A recurring concern was the variability in audio fidelity. Participants involved in broadcast or radio drama production indicated that the AI-generated sounds sometimes lacked the polish of BBC-SFX-UI. For instance, P-$16$ commented, \textit{``The output of the AI-assisted Sound Searching System has poorer audio quality and may not be as useful as the BBC Sound Effect Library for radio drama tasks.''} However, it is essential to note that this limitation is not inherent to the CLAP-UI system itself but rather stems from the quality of the AudioSet samples used in this study. The system is compatible with higher-fidelity datasets, and its performance could be further enhanced with improved audio data.

Citing a lack of written descriptions as a major drawback, P-$3$~(Q1: $2$, Q2: $5$) rated the potential future usage of the AI system with a low score of $2$, as text descriptions are the primary reference for selecting sound effects. They remarked \textit{``often I select effects to audition based on the written description. The AI-assisted search engine didn't have this.''}
At the same time, P-$3$ gives a score of $5$ for Q2 and notes that \textit{``I use a search catalog of a local FX drive daily to find sound effects. The BBC SFX search is always too slow! I have to work extremely quickly.''}. This indicates that both CLAP-UI and BBC-SFX-UI were equally unimpressive for their needs. This highlights the importance of system responsiveness and the importance of textual descriptions in sound-searching workflows. 

% The absence of descriptive metadata in the AI system was also cited as a drawback.
Participants also highlighted that written descriptions in BBC-SFX-UI facilitated quicker selection, as they could filter sounds without listening to each one. P-$2$ remarked, \textit{``I didn’t like the fact that there weren’t any descriptions (as there are in the BBC system), which meant you couldn’t discount a sound without actually listening to it.''} This sentiment was echoed by P-$6$, who suggested, \textit{``It would be better if the search results also come with a text description like the BBC platform. That helps with our usage even faster.''} This limitation could potentially be alleviated with audio captioning systems~\cite{liu2022leveraging, mei2023wavcaps}, which will be part of our future studies.

Furthermore, AI-assisted sound searching may not suit all scenarios. For example, P-$1$ (Q1: $5$, Q2: $3$) commented \textit{``As an addition to the sound library, it’s fine. However, for my use, I need specific sounds—e.g., bird calls or locations around the world. I can't see a scenario where we'd use AI sound for those.''}
Since the CLAP-UI system cannot specify the geographic location of target sounds, it may not be a helpful tool in such cases.

% Overall, while participants generally appreciated the CLAP-UI system’s usability and relevance, specific limitations, such as responsiveness, the lack of textual descriptions, and its inability to handle certain specialized queries, were noted as areas for improvement.

Another drawback was the inconsistent handling of complex scripts. While CLAP-UI excelled at simple queries, multi-element scripts often posed challenges. P-$5$ observed that \textit{``When fusing more than 2 sound requirements, the AI system always ignore one requirement. Like for the jazz piano + footstep, it generates excellent jazz piano but totally ignores the footsteps.''} P-$14$ echoed similar concerns, noting that both systems struggled with scenarios involving combinations of distinct elements. 
This issue may stem from limitations in the search algorithm but may also be because the specific sound effect combinations are originally absent in the dataset. However, as the CLAP-UI system is easily scalable to larger datasets, it holds greater potential for covering a broader spectrum of sound scenes in the future.

Lastly, some users observed redundancies in the results returned by CLAP-UI. For instance, P-$3$ remarked, \textit{``In one example, it offered me the same sound effect multiple times. The variety of suggestions was too limited.''} This highlights the need for de-duplication~\cite{deduplication2014} as a potentially important step in CLAP-UI to minimize repeated results and enhance the diversity of suggestions.

% \subsubsection{Summary of Performance and Usage Patterns}

% Overall, frequent users of sound libraries (e.g., daily or weekly) were more likely to praise the AI-assisted system for its relevance and efficiency, with many noting its potential to streamline daily workflows. Participant~7, who uses sound libraries several times a week, remarked, \textit{“The AI system can … find out the audio that includes most of the elements in the prompt.”} However, even occasional users, such as Participant~19, showed enthusiasm for the system, stating, \textit{“It far surpasses the BBC Sound Effect Library.”} On the other hand, rare users like Participant~1 were more critical, focusing on specific limitations in audio quality and prompt handling.

These findings suggest that while CLAP-UI demonstrates considerable potential in enhancing the efficiency and creativity of sound retrieval workflows, further refinements in metadata, user interface, data curation, and complex query handling are recommended to maximize its potential. 

\begin{figure}[t!]
    \centering % Center the figure on the page
    \includegraphics[width=1.0\linewidth]{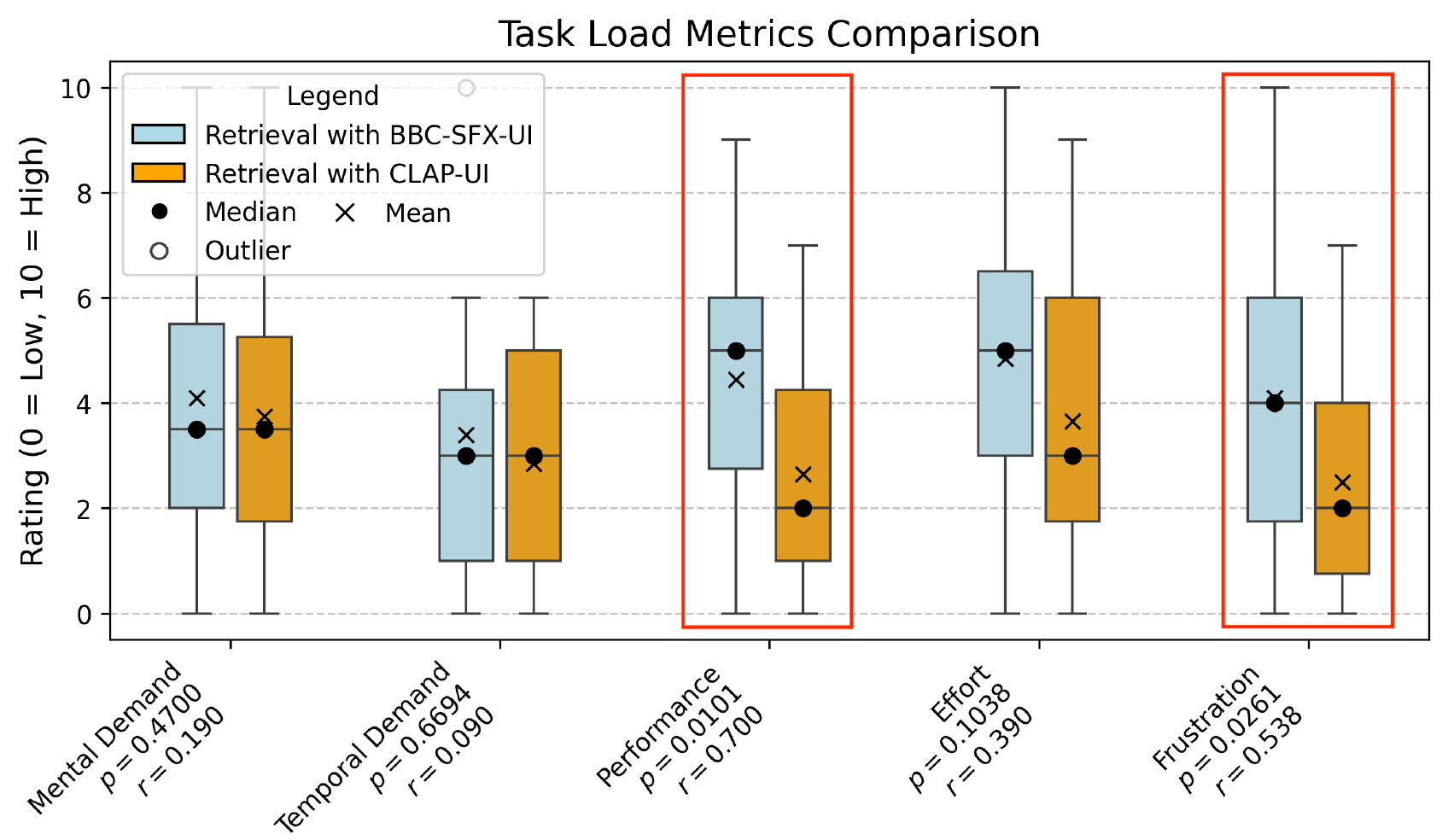}
    \caption{Task load index evaluation result.}
    \vspace{-1em}
    \label{fig:nasa_evaluation} % Optional: Add a label for referencing
\end{figure}

\subsection{Workload Analysis}

The workload experienced by participants during the Sound Source Retrieval task was evaluated using the dimensions defined in the modified NASA-TLX~\cite{hart2006nasa}, encompassing five dimensions: mental demand, temporal demand, performance, effort, and frustration, which result is shown in Figure~\ref{fig:nasa_evaluation}.

% \begin{table}[htbp]
%     \centering
%     \caption{Wilcoxon Test Results for Various Metrics}
%     \label{tab:wilcoxon_results}
%     \begin{tabular}{lcc}
%         \hline
%         \textbf{Metric} & \textbf{Wilcoxon Test Statistic} & \textbf{Wilcoxon p-value} \\ \hline
%         Mental Demand & $47.5$ & $0.470$ \\ \hline
%         Temporal Demand & $39.5$ & $0.669$ \\ \hline
%         Performance & $18.5$ & $\mathbf{0.010}$ \\ \hline
%         Effort & $31.5$ & $0.104$ \\ \hline
%         Frustration & $21.0$ & $\mathbf{0.026}$ \\ \hline
%     \end{tabular}
% \end{table}

% Quantitative results from Wilcoxon signed-rank tests, as shown in Table~\ref{tab:wilcoxon_results}, reveal nuanced differences between the AI-assisted system (CLAP-UI) and the BBC-SFX library.

\subsubsection{Mental and Temporal Demands}

No significant differences were observed in mental demand (\(p = 0.470\)) and temporal demand (\(p = 0.669\)) between CLAP-UI and BBC-SFX-UI, indicating that participants perceived the cognitive complexity and time pressure of tasks to be comparable. 
% Despite the statistical equivalence, several participants noted qualitative differences. For example, Participant~8 highlighted, \textit{``I felt I could get there much faster with the AI-assisted system,''} suggesting that the intuitive query structure of CLAP-UI might have offset its novelty, balancing any initial cognitive overhead.

\subsubsection{Performance and Effort}

Performance ratings showed a significant improvement for CLAP-UI over BBC-SFX-UI (\(p = 0.010, r=0.700\)), with participants consistently reporting that CLAP-UI provided more accurate and relevant results. This improvement aligns with qualitative insights, such as P-$10$’s observation that \textit{``the BBC library often gave irrelevant results, while CLAP-UI provided more useful options.''} While effort did not achieve statistical significance (\(p = 0.1038, r=0.390\)), a trend toward reduced effort with CLAP-UI was evident. Participants attributed this to the system’s ability to retrieve results efficiently with minimal re-querying.
% , as noted by P-$9$: \textit{``I can generate multiple audio clips from similar scenes quickly and efficiently.''}

\subsubsection{Frustration}

The frustration dimension showed a significant reduction with CLAP-UI compared to BBC-SFX-UI (\(p = 0.026, r=0.538\)). Participants frequently cited frustration with the BBC-SFX-UI due to irrelevant results and redundant outputs, whereas CLAP-UI’s semantic relevance and natural language interface alleviated these issues. 
% Participant~19 remarked, \textit{``The AI-assisted system always worked for me, reducing the annoyance of manually searching for sound effects,''} highlighting the potential of CLAP-UI to streamline workflows and improve user experience.

% \subsubsection{Implications for Workload Distribution}
The findings on the task load index indicate that CLAP-UI can alleviate specific aspects of workload, such as frustration and perceived performance while maintaining similar mental and temporal demands as the traditional BBC-SFX-UI. This balance suggests that CLAP-UI is effective at integrating into existing workflows without introducing significant cognitive burdens.
% making it a viable alternative for professional sound retrieval tasks.

% Analysis of any other feedback

\section{Discussions and Limitations}

While this study highlights the benefits and potential of the CLAP-based sound searching system (CLAP-UI), several limitations must be acknowledged. First, the AudioSet dataset used in CLAP-UI may have influenced participant evaluations. Although extensive, AudioSet contains lower-fidelity audio compared to the BBC SFX, which may have affected user perceptions of audio quality. This limitation, however, is not inherent to the CLAP-UI system itself, as it can be adapted to work with high-quality datasets.

Second, the CLAP-UI system is not a fully developed product and lacks the refinement of systems created by professional web developers. For instance, the user interface may have influenced the user experience, as it lacks the polish and intuitiveness of more established systems. Additionally, the absence of descriptive metadata in CLAP-UI posed significant challenges for participants, as they were unable to evaluate results without manually listening to each one. Another challenge arises from the fundamental shift in search logic employed by CLAP-UI, moving away from traditional word-matching methods to a more semantic, natural-language-based approach. This change created some confusion among participants, as they often expected the AI-assisted system to do everything, such as handle complex scenes without requiring breakdowns into simpler components. For example, Participant~1 remarked \textit{``With both AI and the sound library, I'd still be breaking down the sound search into its components ... as the timing and levels of those are specific to the script and story.''} To improve user experience, it is essential to provide clear guidance on how the AI sound-searching system functions and what users can reasonably expect. 

Third, CLAP-UI occasionally does not work well on scripts involving multiple elements (e.g., \textit{“jazz piano with footsteps”}), leading to partial fulfillment of user queries. While this limitation may stem from the inherent challenges in modeling such complexity in CLAP-based systems, it also reflects potential dataset coverage issues. Addressing these challenges with a larger dataset scale is an important direction for future research. 

Another limitation of the CLAP-UI system we found is the lack of explainability in its search process. Several participants reported difficulty in understanding why certain queries failed to produce satisfactory results or which aspects of their prompts needed adjustment. For instance, when a sound could not be located, users were left uncertain whether the issue stemmed from the dataset limitations, the system retrieval logic, or the phrasing of their input. This ambiguity and lack of explainability made it challenging for participants to refine their queries. 

Finally, the study's participant pool, though diverse, is limited in size. While the participants included professionals with relevant expertise, further studies with a broader audience would provide deeper insights into the scalability and robustness of CLAP-UI in various use cases.

\section{Conclusions}

This study compared a CLAP-based sound searching system (CLAP-UI) with the word-matching-based sound searching system implemented in the BBC Sound Effects Library (BBC-SFX-UI). Our result demonstrates that CLAP-UI offers significant advantages in performance, frustration reduction, and creative flexibility, largely due to its natural language querying and better semantic relevance. However, limitations such as reliance on lower fidelity datasets, lack of metadata, and challenges in handling complex prompts highlight areas for improvement. Despite these limitations, CLAP-UI represents a promising advancement in sound retrieval technologies, with the potential to streamline workflows, reduce cognitive demands, and inspire creativity in audio production. Future work should address these limitations by enhancing metadata integration, improving query modeling, and leveraging higher-quality datasets.

\ifthenelse{\equal{\Publicationready}{true}}
  {
\section{Acknowledgments}  This research was partly supported by the British Broadcasting Corporation Research and Development~(BBC R\&D), Engineering and Physical Sciences Research Council (EPSRC) Grant EP/T019751/1 ``AI for Sound'', and a PhD scholarship from the Centre for Vision, Speech and Signal Processing (CVSSP), Faculty of Engineering and Physical Science (FEPS), University of Surrey. For the purpose of open access, the authors have applied a Creative Commons Attribution (CC BY) license to any Author Accepted Manuscript version arising.}
  {}
  
% \begin{figure}
%     \centering
%     \includegraphics[width=1.0\linewidth]{figures/UI-old.png}
%     \caption{Caption}
%     \label{fig:enter-label}
% \end{figure}

\bibliography{ref}
\bibliographystyle{IEEEtran}

\end{document}